\documentclass[letterpaper,twocolumn,10pt]{article}

\usepackage[latin1]{inputenc}
\usepackage[english]{babel}
\usepackage{scs}
\usepackage{graphicx}
\usepackage{caption}
\usepackage{cite}
\usepackage{subcaption}
\usepackage{amssymb,amsmath}
\usepackage{hyperref}
\usepackage[vlined,boxruled]{algorithm2e}

\SetAlFnt{\small}
\SetAlCapFnt{\small}
\SetAlCapNameFnt{\small}

\begin{document}

\title{Stochastic Agent-Based Simulations of Social Networks}

\author{
	Garrett Bernstein and Kyle O'Brien  \\
	MIT Lincoln Laboratory; 244 Wood Street; Lexington, MA 02421 \\
	(\href{mailto:garrett.bernstein@ll.mit.edu}{garrett.bernstein}, \href{mailto:kyle.obrien@ll.mit.edu}{kyle.obrien})@ll.mit.edu
}

\maketitle


\begin{abstract}

	\noindent
	The rapidly growing field of network analytics requires data sets for use in evaluation. Real world data often lack truth and simulated data lack narrative fidelity or statistical generality. This paper presents a novel, mixed-membership, agent-based simulation model to generate activity data with narrative power while providing statistical diversity through random draws. The model generalizes to a variety of network activity types such as Internet and cellular communications, human mobility, and social network interactions. The simulated actions over all agents can then drive an application specific observational model to render measurements as one would collect in real-world experiments. We apply this framework to human mobility and demonstrate its utility in generating high fidelity traffic data for network analytics. \footnote{This work is sponsored by the Assistant Secretary of Defense for Research \& Engineering under Air Force Contract \#FA8721-05-C-0002.  Opinions, interpretations, conclusions and recommendations are those of the author and are not necessarily endorsed by the United States Government}

\end{abstract}

\section{INTRODUCTION} 
 
	Understanding phenomena in real world networks is a prominent field of research in many areas. There are a wide variety of inferential tasks on phenomena in communication, social, and biological networks; for example, email traffic between employees of a company \cite{enron}, vehicle traffic between physical locations \cite{network_wami}, collaborations between scientists \cite{scientists}, protein-protein interactions \cite{proteins}. These studies range from clustering nodes into discrete communities to anomaly detection to inferring attributes on individual nodes to searching for specific activity embedded in background population clutter. The analytical approaches taken to study networks require vast amounts of complex, truthed data for algorithmic verification and there is currently a dearth of sufficient network data. We explore the causes of this data problem and introduce our solution: a novel, two-tiered model that addresses drawbacks in current options. The first tier is an agent-based, mixed-membership, activity model that is easy to parametrize and abstractly generates agents' actions over time. The second tier is an application specific observational model that supplies researchers with the simulated sensor data necessary to conduct experiments.
	
	Researchers face a trilemma of inadequate data from real world datasets, statistical simulation models, and agent-based simulation models. Large-scale real world data sets are expensive to collect and difficult to obtain high fidelity ground truth for. Statistical models, such as Er\H{o}ds-R\'{e}nyi, Chung-Lu, and blockmodels, have parameters that are easy to specify and allow for simple replication of large-scale data sets. What is often missing, however, is the ability to encode narratives into the data because there is no sense of individual agents, just interactions between nodes. Hand-crafted agent-based models address this problem by allowing for narratives in the sense of specific actions taken by an agent throughout time. Those networks, however, may not result in the desired aggregate statistical behavior and are usually difficult to adapt to other applications.
	
\begin{figure}[ht]
		\centering
		\includegraphics[width=.25\textwidth]{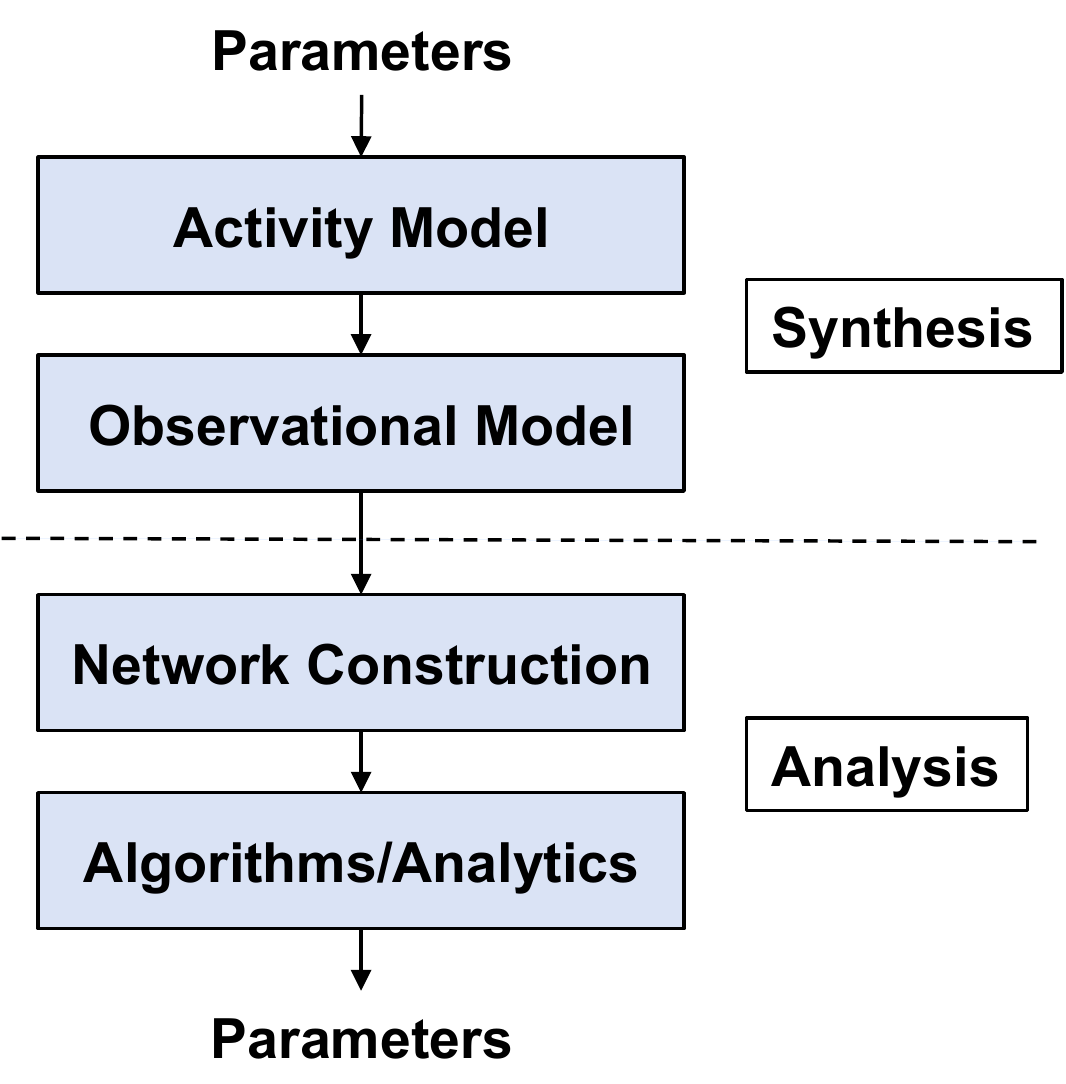}
		\caption{Network analytics workflow}
		\label{fig:flow_chart}
	\end{figure}

	Additionally, generating network data is only half of the modeling problem. In the real world, data sets are not delivered as clean networks with nodes and edges. Instead, algorithms must process them in the form of noisy sensor observations. Therefore, simply using an activity model is not enough to effectively simulate data for network analytics. Instead, the simulation must be augmented with an observational model for the particular application we wish to study. Once the observable sensor data has been simulated, we can then feed it into the desired network analytics and construct a network. This flow can be seen in Figure \ref{fig:flow_chart} where the top half depicts the data synthesis aspect, with parameters describing a population's behavior, the activity model generating the population's actions based on the parameters, and the observational transforming the actions into observable sensor data. The bottom half depicts the network analysis problem, with networks being constructed from observed sensor data and then algorithms inferring desired properties and parameters of the network.
	
 	This paper is organized as follows: In Section \ref{sec:background} we give a brief background of previous work on simulation models and discuss their advantages and disadvantages. In Section \ref{sec:activity_model} we introduce our activity model that employs the most desirable aspects of statistical and agent-based models. This model uses high-level population parameters to drive an agent-based narrative, enabling it to create rich network datasets but also allows for generation of numerous, statistically similar datasets for Monte Carlo purposes in analyzing network algorithms. In Section \ref{sec:observational_model} we further the utility of the activity model by introducing the observational model that takes abstract network interactions and transforms them into the simulated output of a real sensor, allowing for realistic experimentation methods. We conclude in Section \ref{sec:conclusion} and discuss future work.


\section{BACKGROUND} 
\label{sec:background}

	Network data come from two sources: collected real world data and simulated data. Real world data sets are are exactly on what the inferential tasks will be run in deployment and thus can claim the highest fidelity. In practice, however, collection of this type of data faces many hurdles. Privacy of personal data can cause both regulatory issues and hinder avenues of potential research, such as \cite{pentland_coevolution} needing to follow experimental oversight regulations on cell phone GPS tracking. The desired data can take a long time to collect and only result in one data set, which is insufficient for Monte Carlo purposes. For example, \cite{karate} required two years to observe a social network of only thirty-four people. Possibly most importantly for algorithm development, collecting sufficiently representative and comprehensive data on a large scale is a daunting task. \cite{xie} explores the prediction of clustering in protein interactions collected by \cite{gavin}, but that data set ``represents only a snapshot of the proteome averaged over all phases of the cell cycle.''
	
	Simulation models can be a powerful alternative for obtaining the requisite experimental data and can be broken into two categories. First, statistical models employ high-level statistics to describe the aggregate behavior of a population. This allows researchers to closely match the simulated population to the desired real world population, but generally leads to network interactions that do not have narrative fidelity. The simplest, such as Erd\H{o}s-R\'{e}yni or Chung-Lu , are easy to parameterize and can quickly provide many large iterations but are only able to create populations with homogeneous behavior. \cite{erdos-reyni} \cite{chung-lu} Slightly more complicated models, such as RMAT \cite{rmat} and Blockmodels \cite{mmsb}, address that issue by generating populations with power-law degree distributions and with block community structures, respectively, properties which are prevalent in many real world populations. These models, however, only specify interactions between nodes and thus fail to encode a narrative of specific actions over time that accurately reflect the constraints and behavior of the target data.
	
	The second type of simulation model, agent-based, fixes the lack of narrative by building the network from the ground up. Instantiating individual agents and directly simulating their behavior necessitates that the resulting network obeys the individual constraints imposed on agents' activities. Achieving a high-fidelity narrative comes at a cost, however, as agent-based models struggle with some aspects that come naturally to statistical models. A simulated vehicle motion dataset made available by the National Geospatial-Intelligence Agency (NGA) is a one-off, hand-crafted, agent-based model that simulates over 4000 people driving between over 5000 locations throughout Baghdad, Iraq over a 48-hour time period. Because the network was carefully handcrafted it reflects real world behavior relatively well but it took 2-man years to create and only one instance of it exists, so it cannot be used for Monte Carlo experiments.
	
	Agent-based models tend to be very specific in their application focus and require highly tuned parameters and intricate understanding of the application space. \cite{biowar} introduces a powerful agent-based simulation model which provides the necessary data to successfully study the spread of disease through a city in the context of bioterrorism. The model employs pertinent data sources, such as that from census, school districts, drug purchases, emergency room visits, etc. The attention to detail ensures a realistic simulation model but necessitates expert knowledge and copious amounts of target data, thus making it difficult to transfer the model to other desirable applications or even for someone not fully experienced with the model to re-parameterize it.
	


\section{ACTIVITY MODEL} 
\label{sec:activity_model}
	
	As mentioned in the Introduction, the development of network inference algorithms suffers from a lack of truthed, high-fidelity data. Real data is hard to collect and truthing it often runs into privacy concerns. Simulated data can be generated but popular generation methods lack statistical fidelity or narrative flow. In this section we introduce a mixed-membership, agent-based model that aims to provide both desirable aggregate statistics and realistic agent interactions.
	
	Instead of directly simulating nodes and edges our approach to simulating network data recognizes the fact that most network data are created by individuals taking actions over time. This could be users clicking on websites, people sending emails, or vehicles driving to locations. We leverage this insight by directly simulating agents' actions over time.

	Just simulating agents' actions over time would be difficult both to parameterize and to allow agents to be diverse in their nature. To add richness to the data we introduce the concept of roles, which we define as the agent's intention in executing an action. In this way, prior to selecting an action the agent first decides what role it will adopt and then chooses an action based on that role.
	
	The model draws from the widely-used concept of mixed-membership in which data are treated as a mixture of classes. Latent Dirichlet Allocation \cite{lda} treats documents as a mixture of thematic topics. Optical Character Recognition \cite{bishop} can be implemented to treat a written character as a mixture of potential digits. Mixed-Membership Stochastic Blockmodels \cite{mmsb} treat networks as a mixture of community membership. We extend these concepts so that instead of agents taking on only one role throughout the entire simulation they can instead have a mixture of intentions at any given time

	To add further fidelity to the data we enable agents to dynamically change their mixture of roles over the total time of the simulation. The mixture of roles can easily be made cyclical to reflect diurnal patterns of behavior.
	
	\subsection{Mathematical Description} 
	\label{sub:plate_model}
	
		The essence of the the activity model can be broken into three parts for each event of choosing an action: drawing the time at which the agent's event occurs, drawing the role the agent adopts during the event, and drawing the action the agent actually executes as a result of the event.
		
\begin{figure}[ht]
			\centering
			\includegraphics[width=.4\textwidth]{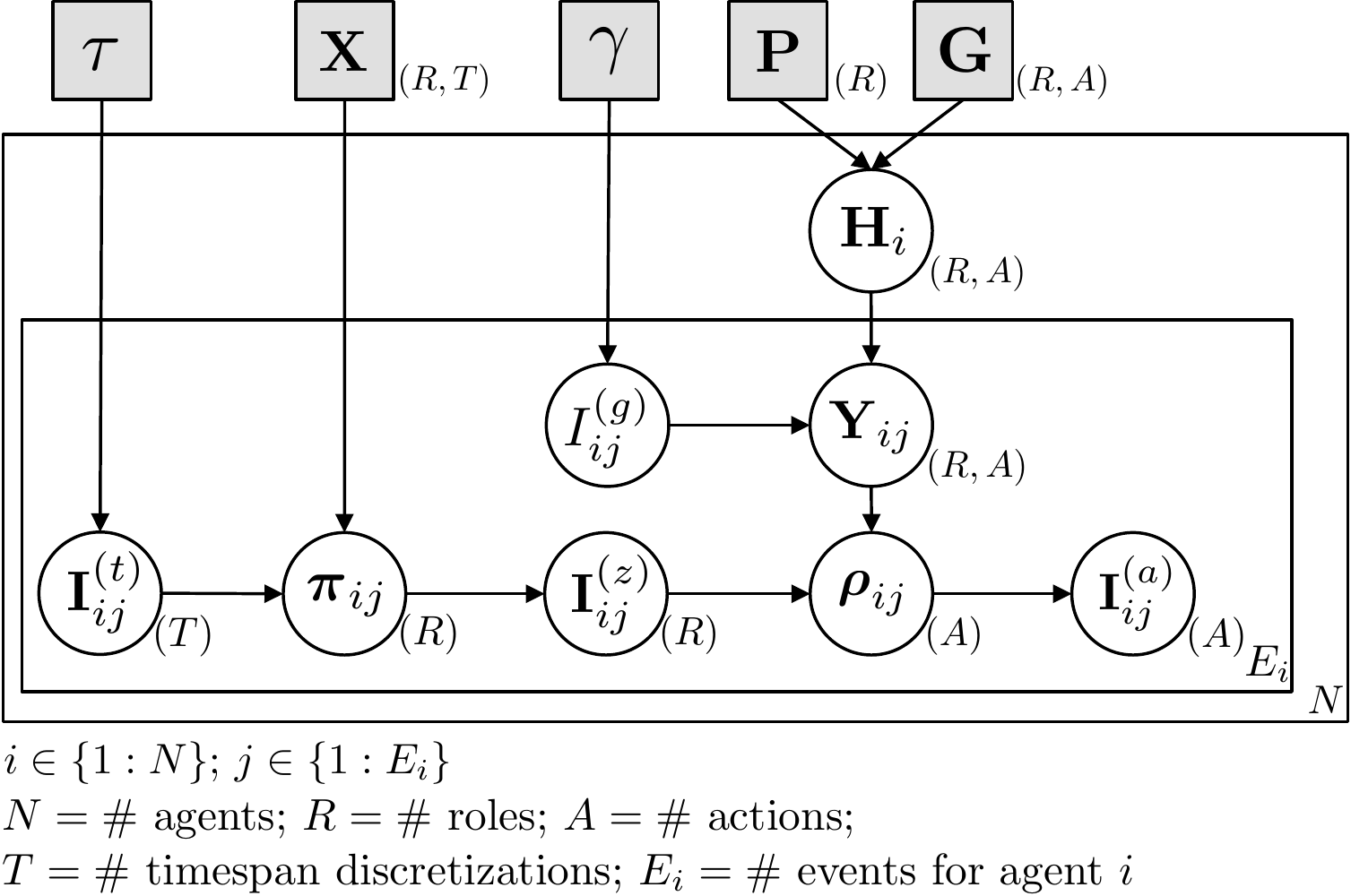}
			\caption{Activity model}
			\label{fig:plate_model}
		\end{figure}

		The plate model in Figure \ref{fig:plate_model} represents the activity model that will be described in detail in this section. Plate models are convenient methods to depict algorithms with replicated variables. Shaded boxes denote input parameters, circles denote random variables, arrows denote variable dependencies, and plates denote repeated variables, with the variable in the bottom corner of the plate denoting the number of replications.  We let $N$ be the number of agents, $R$ be the number of roles, $A$ be the number of possible actions, $T$ be the number of timespans in which we discretize the total time of the simulation (e.g. day, evening, night), and $E_i$ be the number of events for the $i^\text{th}$ agent
		
		The algorithm depicted by the plate model in Figure \ref{fig:plate_model} is written out in Algorithm \ref{fig:activity_model_pseudocode}.

			\begin{algorithm}[ht]
		 		\SetAlgoLined

		 		\KwData{User defines $\tau$, $\mu$, $\mathbf{X}$, $\gamma$, $\mathbf{P}$, $\mathbf{G}$}

		 		\KwResult{A list of actions $a_i$ for each agent $i$ and the time of each action}		

				\ForEach{agent $i \in \{1:N\}$}{

					\begin{align*}
						&\text{$\backslash\backslash\oslash$ denotes element-wise division}\\
						&\mathbf{H}_i \sim \text{Multinomial}(G \oslash G \mathbf{1}_{(A)},P) \\
						&E_i \sim \text{Poisson}(\mu\tau)
					\end{align*}

					\For{event $j \in \{1:E_i\}$}{

						\begin{align*}
							\mathbf{I}^{(t)}_{ij} &\sim \text{Uniform}(0,\tau) \\
							\pi_{ij} &\sim \text{Dirichlet}(\mathbf{I}^{(t)}_{ij}X) \\
							\mathbf{I}^{(z)}_{ij} &\sim \text{Multinomial}(\mathbf{\pi}_{ij}) \\
							I^{(g)}_{ij} &\sim \text{Bernoulli}(\gamma) \\
							\mathbf{Y}_{ij} &= I^{(g)}_{ij} \mathbf{H}_i + (1 - I^{(g)}_{ij}) (\neg \mathbf{H}_i) \\
							\mathbf{\rho}_{ij} &\sim \text{Dirichlet}(\mathbf{I}^{(z)}_{ij}\mathbf{Y}_{ij}) \\
							\mathbf{a}_{ij} &\sim \text{Multinomial}(\mathbf{\rho}_{ij}) 
						\end{align*}
					}
	 			}

				\caption{Activity model algorithm}
				\label{fig:activity_model_pseudocode}
			\end{algorithm}
		
		\subsubsection*{Choosing the event's time}
			
			Choosing the times of each agent's events first requires deciding how many events will occur for the agent. We draw the number of events $E_i \sim \text{Poisson}(\mu\tau)$, where the input $\tau$ is the total time of the simulation and the input $\mu$ is the average amount of time before the agent waits to initiate another event. The times of the $E_i$ events are then drawn uniformly over the entire time of the simulation.
			
			To lend fidelity to agent behavior we allow their parameterizations to take on different values in the input $T$ timespans (e.g. propensity towards work during the day, restaurants in the evening, and home at night).  We create the indicator $\mathbf{I^{(t)}_{ij}}$, which specifies during which of the $T$ timespans the $j^\text{th}$ event for the $i^\text{th}$ agent occurs.
			
			In actual implementation we employ Poisson count-time duality to determine the number and timing of events by having each agent draw sequential waiting times between actions from $\sim \text{Exponential}(\mu)$. This allows us to build action durations into the wait times and ensures a realistic narrative in which an agent's actions cannot overlap.
			
		\subsubsection*{Choosing the agent's role}
			
			Once the agent has picked a time for each of its events it then chooses roles for those actions. We wish to avoid inputting a specific distribution over roles for each agent as this will lead to bias in Monte Carlo experiments. Instead we input $\mathbf{X}$,  the Dirichlet concentration parameters, which specifies the propensity of the population toward roles at each timespan. To obtain the actual distribution over roles for an event we draw $\mathbf{\pi}_{ij} \sim \text{Dirichlet}(\mathbf{I}^{(t)}_{ij} \mathbf{X})$. The actual role for that action is then drawn as the indicator $\mathbf{I}^{(z)}_{ij} \sim \text{Multinomial}(\boldsymbol\pi_{ij},1)$.
			
		\subsubsection*{Choosing the agent's action}
			
			Once the agent has drawn a role for its $j^\text{th}$ event it then must draw an actual action to execute. Each action belongs to one of the roles, as specified by the user input $\mathbf{G}$, and can only be executed by an agent in that role. Again, to avoid bias, instead of inputting the distribution over all actions we will draw the distribution as $\boldsymbol\rho_{ij} \sim \text{Dirichlet}(\mathbf{I}^{(z)}_{ij} \mathbf{Y}_{ij})$. $\mathbf{Y}_{ij}$ is the Dirichlet concentration parameters which specify the propensity of an agent towards every action given the role of the event.
			
			 $\mathbf{Y}_{ij}$ is constructed in two conceptual parts, whether or not this is a normal event and which are the normal actions. At the beginning of the simulation, every agent must determine which actions it deems normal and which actions are abnormal. The number of each type of action the agent deems normal is specified by the user input $\mathbf{P}$. In splitting the actions this way the agent will have a routine over time but still be allowed deviate from the norm. The agent draws its normal actions from $\mathbf{H}_i \sim \text{Multinomial}(\mathbf{G} \oslash \mathbf{G} \mathbf{1}_{(A)},\mathbf{P})$, where $\oslash$ denotes element-wise division. This multinomial draw is set up to uniformly draw normal actions from all possible actions of each type.
			 
			 To determine whether each event is normal or abnormal the agent draws the indicator $I^{(g)}_{ij} \sim \text{Bernoulli}(\gamma)$. Then the agent constructs its Dirichlet propensity towards all actions for that event by $\mathbf{Y}_{ij} = \mathbf{I}^{(g)}_{ij} \mathbf{H}_i + (1 - \mathbf{I}^{(g)}_{ij}) (\neg \mathbf{H}_i)$, where $\neg$ denotes negation. From that the agent draws its probability distribution over all actions as $\boldsymbol\rho_{ij} \sim \text{Dirichlet}(\mathbf{I}^{(z)}_{ij} \mathbf{Y}_{ij})$. Finally, the agent draws the actual action executed for that event as the indicator $\mathbf{I}^{(a)}_{ij} \sim \text{Multinomial}(\boldsymbol\rho_{ij},1)$.		
				

	\subsection{Activity Model Results} 
	\label{sub:activity_model_results}
	
		We present two types of results to show the capabilities of the activity model. First we display the actions of an individual agent under different parameter settings to show the richness of model. Second, we tune the parameters to a target data set to demonstrate the ability to simulate data with specific desired behavior and thus successfully provide sufficient data for network analytic experiments.
		
		So far we have discussed the activity model in an application-agnostic manner. To ground the discussion of the results we will place the activity model in the human mobility modality, specifically people driving throughout a city. This application and the motivation for network analysis of human mobility will be more fully discussed in Section \ref{sec:observational_model} but for now it suffices to say that agents are people, an action is a person driving from their current location to a destination location, and a role is the person's intention to drive to the destination. For example, a person may assume a work role and decide to drive to their company's building. 
		
		\subsubsection*{Richness of parameter settings}
		
\begin{figure}[ht]
					\begin{subfigure}[b]{0.3\textwidth}
						\centering
						\includegraphics[width=\textwidth]{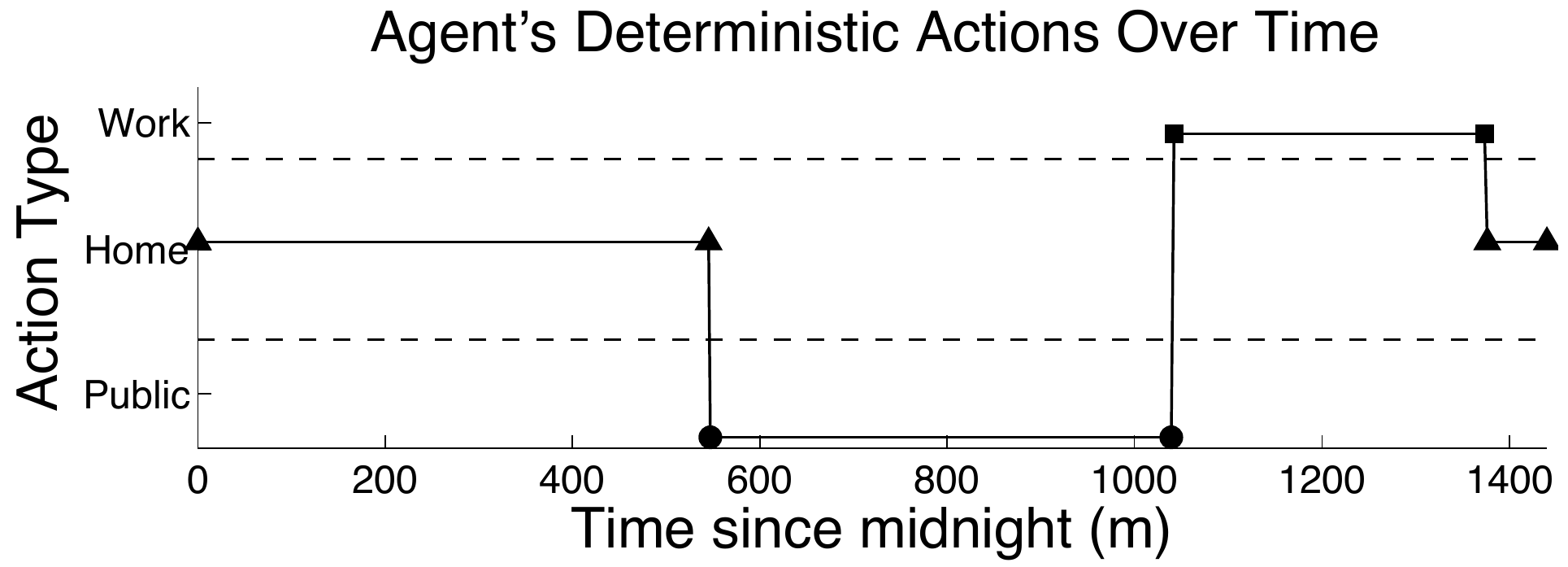}
						\caption{Deterministic parameter settings}
						\label{fig:sanity_check_deterministic}
		       		 \end{subfigure}
			
					\begin{subfigure}[b]{.3\textwidth}
						\centering
						\includegraphics[width=\textwidth]{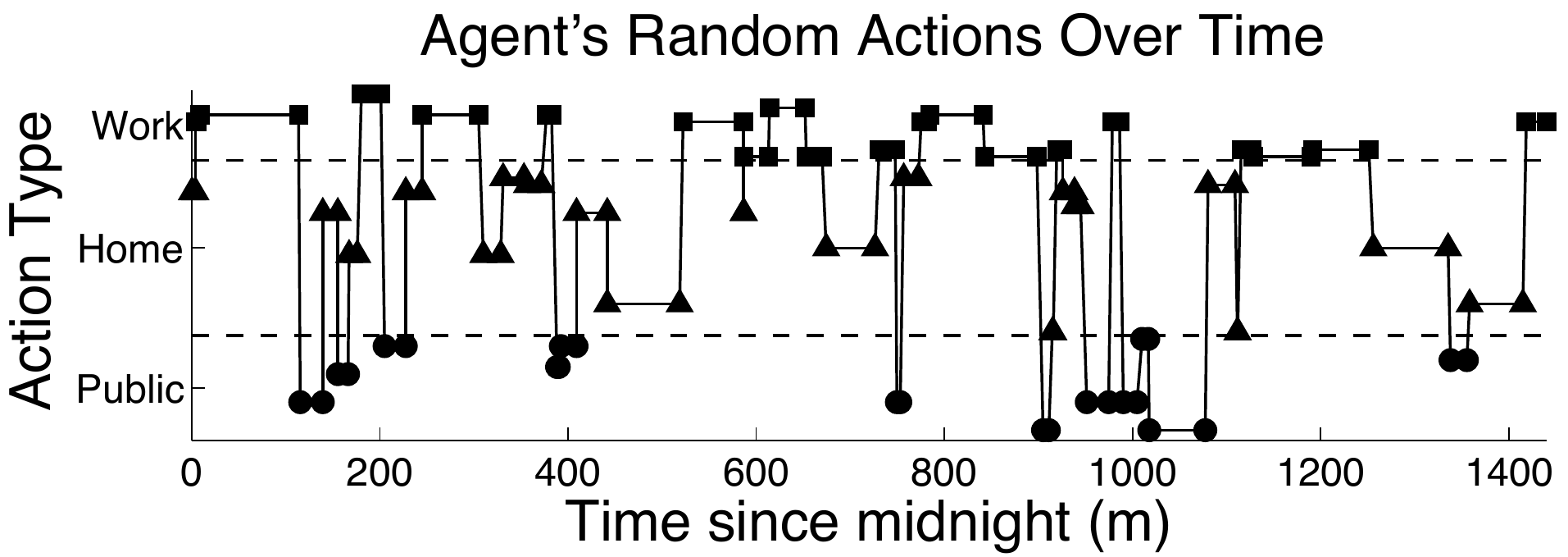}
						\caption{Random parameter settings}
						\label{fig:sanity_check_random}
		     		\end{subfigure}
			
					\begin{subfigure}[b]{0.3\textwidth}
						\centering
						\includegraphics[width=\textwidth]{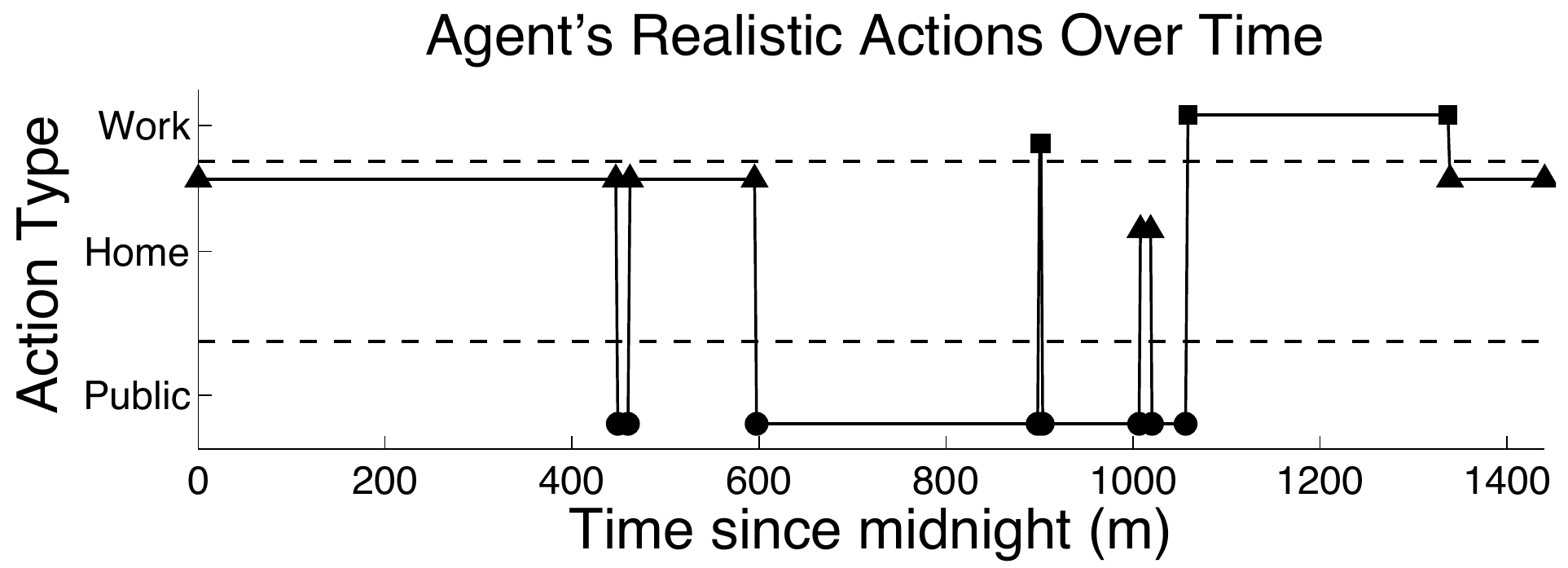}
						\caption{Realistic parameter settings}
						\label{fig:sanity_check_realistic}
	        		\end{subfigure}
					
					\caption{Spatio-temporal plots depicting an agent's actions over a 24-hour time period in a toy example. The $x$-axis is minutes since midnight. The $y$-axis is divided into three roles, with different $y$ values representing different actions and the marker shape also denoting to which role the action belongs.}
					\label{fig:sanity_check}
				\end{figure}

			We show the spectrum of the attainable richness of data with varying parameter settings using a toy human mobility example. We simulate 100 agents taking on three possible roles of Home, Work, and Public, with 25, 5, and 10 locations of each type, respectively. A public location may be a park or sports arena. Figure \ref{fig:sanity_check} provides spatio-temporal plots of a single agent's actions over a 24-hour time period to show the effects of varying parameters on behavior. These plots display the progression of an agent's movements over the duration of the simulation. Time proceeds along the positive x-axis in minutes since midnight. Each integer on the positive y-axis represents a location and the locations are separated vertically with dashed lines by category. A solid horizontal line indicates the agent staying at that single location over that span of time. A diagonal line indicates an event occurring in which the agent chooses and then travels to a new location.
			
			We create three different parameter settings: deterministic, random, and realistic. Figure \ref{fig:sanity_check_deterministic} depicts parameters that cause an agent to deterministically adopt the three roles over specified time spans and choose to execute specified actions for each role. Figure \ref{fig:sanity_check_random} depicts parameters that cause an agent to randomly choose a role at any given point in time and to then randomly choose an action to execute given that role. Figure \ref{fig:sanity_check_realistic} depicts parameters in the middle-ground where an agent has a normal lifestyle but still retains the ability to deviate from the norm.
	
		\subsubsection*{Comparison to target dataset}
		
			Simulated data is only useful if it faithfully reflects the target data. Publicly available real-world network data is difficult to employ in this venue given privacy and security concerns so we will compare our model to the NGA dataset. As discussed in Section \ref{sec:background}, the dataset is a one-off, hand-crafted, agent-based model that simulates over 4000 people driving throughout Baghdad, Iraq over a 48-hour time period. The model is not perfect at simulating realistic movements within a city but it is close enough for our purposes to show that we can achieve a high fidelity comparison to a specific target.
			
			First we parameterized the activity model to match the NGA data as closely as possible. One strength of our model lies in using relatively few, easy-to-set parameters. We directly matched simplistic parameters such as the number of agents, $N$, the types of roles, $R$, the number of locations, $A$, and the location categories, $\mathbf{G}$. With trivial analysis we were able to determine two input parameters not directly accessible in the target data: How many of each actions type an agent deems normal, $\mathbf{P}$, and agent propensities towards roles, $\mathbf{X}$.
			
			After using these parameters to simulate an instance of the data we compared our output to the NGA dataset. In our data agents chose to belong to 0.30 fewer locations of each type on average and visited locations of each type 1.98 fewer times on average. Matching agents' actions so closely to the target allows the simulated data to stand in for the NGA dataset in desired Monte Carlo experiments. The current activity model does not account for community structure in the population so we omit any comparison in that domain, though an expansion of the model to include community membership and affinity towards shared locations is certainly feasible.
			
\begin{figure}[ht]
				\begin{subfigure}[b]{0.4\textwidth}
					\centering
					\includegraphics[width=\textwidth]{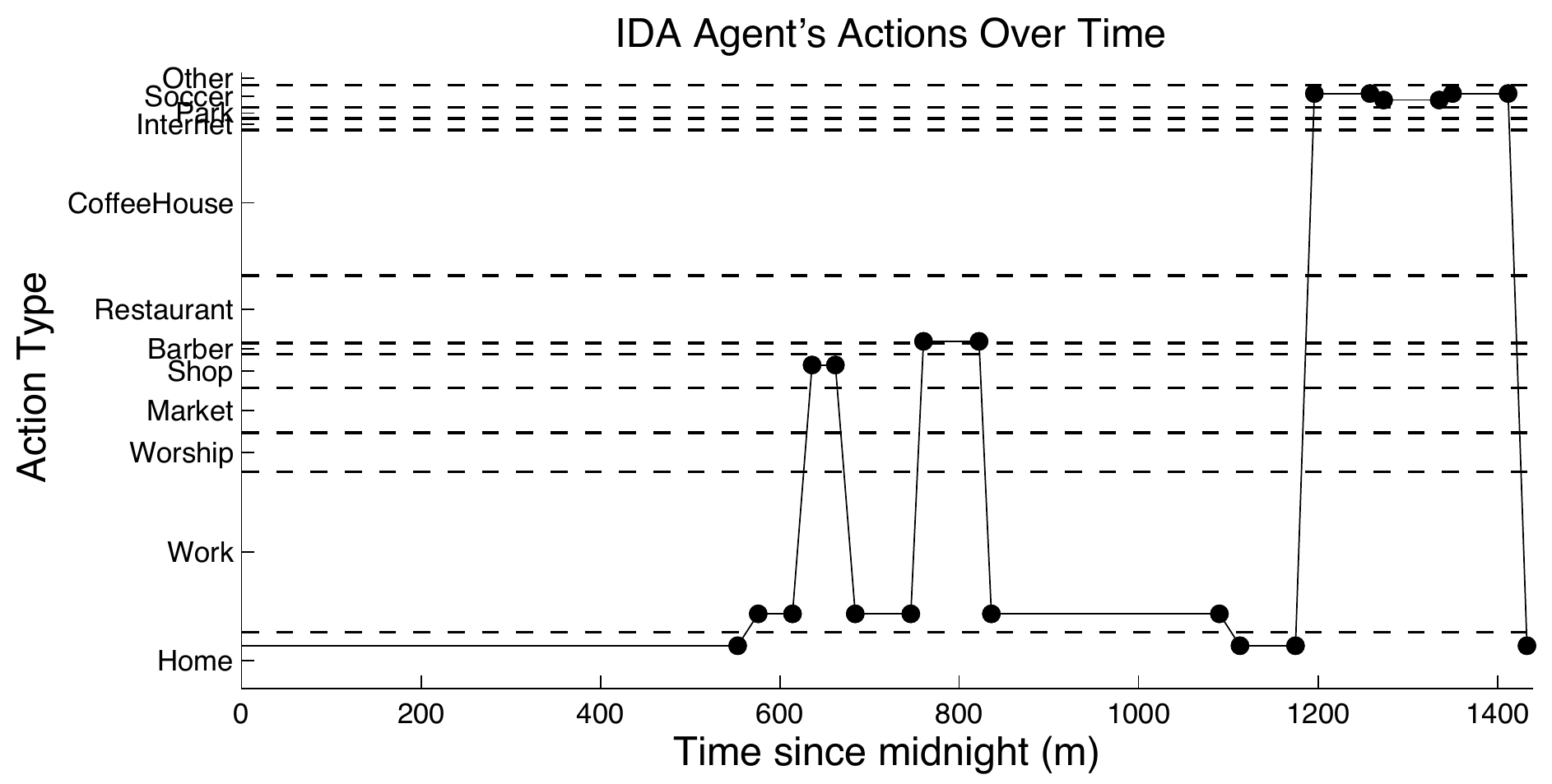}
					\caption{NGA Agent}
					\label{fig:agent_comparison_ida}
	       		 \end{subfigure}
		
				\begin{subfigure}[b]{.4\textwidth}
					\centering
					\includegraphics[width=\textwidth]{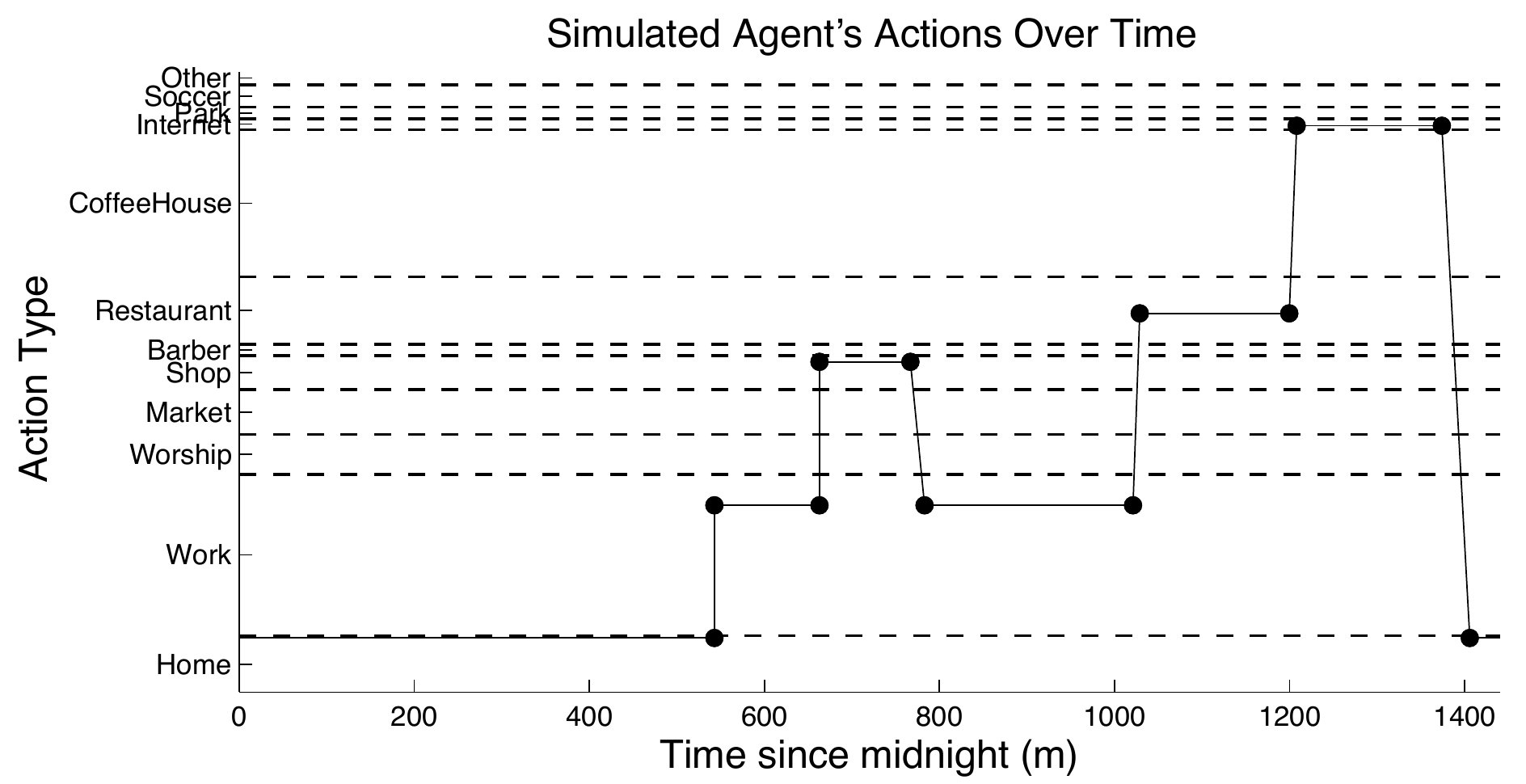}
					\caption{Simulated Agent}
					\label{fig:agent_comparison_simulated}
	     		\end{subfigure}

				\caption{Spatio-temporal plots comparing a simulated agent's and an NGA agent's actions over a 24-hour time period.}
				\label{fig:agent_comparison}
			\end{figure}

			The simulated agent's individual activity also matched that of the NGA agents as can be seen in the two spatio-temporal plots in Figure \ref{fig:agent_comparison}. Over many simulations the agents produce behavior statistically similar but not identical to the target agents, as is desirable for Monte Carlo experiments.
			
			In addition to fidelity, running time is an important consideration in simulating data so that Monte Carlo experiments with large numbers of trials are feasible. To simulate NGA data with 4623 agents and 5444 locations over 24 hours on a standard Mac desktop currently takes approximately 25 minutes on average to draw the actions via the activity model and approximately 35 minutes on average to generate the tracks via the observational model that will be discussed in Section \ref{sec:observational_model}. Both models are parallelizable with either Matlab's Parallel Toolbox or grid computing so we envision being able to significantly speed up the implementation from its current state.



\section{APPLICATION: HUMAN MOBILITY} 
\label{sec:observational_model}

	The previous section introduced a generalized agent-based activity model and then applied the model to a human mobility application by tuning the parameters to simulate the NGA data set. That simulated data contains agents' abstract actions but researchers may also want to study the sensor observations resulting from the execution of those actions.
	
	Therefore, we introduce an observational model for human mobility applications that generates tracks of vehicle movement within a city. The observational model is flexible enough to create simulations for any place in the world for which open-source map data is available. To demonstrate this flexibility and assess the fidelity of the results, we will tailor parameters of the model to match properties of the NGA data set and then show a comparison of the results.

	\subsection{Motivation} 
	\label{sub:motivation}

	Tracking movement from location to location lends itself naturally to network-related data and many interesting inference and prediction applications. For example, in wide-area, aerial surveillance systems, vehicle tracks can be used to forensically reconstruct clandestine terrorist networks. Greater detection can be achieved using network topology modeling to separate the foreground network embedded in a much larger background network \cite{network_wami}. Location data recorded by mobile phones has been used to track the movement of large groups of students on a university campus to learn more about human behavior and social network formation \cite{pentland_coevolution}. Location data from mobile phone activity can also be used to measure spatiotemporal changes in population to infer land use and supplement urban planning and zoning regulations \cite{gonzalez_landuse}.

	Collecting large-scale, persistent data to study human mobility, however, is especially difficult. Aerial surveillance systems are limited by practical issues such as flight endurance limitations and line-of-sight occlusions. Even in the absence of these issues, accurate vehicle tracking is very difficult. Furthermore, ground truth is required to assess the accuracy of the observed network, but activity of interest rarely includes ground truth or requires pre-planned instrumentation such as GPS sensors. While cell phones equipped with GPS sensors are ubiquitous, such data is difficult to obtain because of legal and privacy concerns. Simulating large-scale, persistent data sets to study human mobility would greatly benefit the development of algorithms for applications where such data is limited.
	
		
	\subsection{Observational Model} 
	\label{sub:observational_model}	

		The observational model is comprised of four parts. First, we adapt the activity model to our human mobility application by specifying agents, actions, and roles in terms of human mobility. Next, we obtain map data and define the road network on which we will create vehicle tracks. Then we label nodes in the network according to the type of physical location they represent. The final step is to create vehicle tracks by defining paths between intended locations and modeling sensor observations of that motion.
	
		Our observation model can be made arbitrarily complex to replicate specific sensor observability and noise characteristics. It is also possible to simulate observational data for any conceivable type of sensor. However, detailed sensor modeling is beyond the scope of this paper and we assume that algorithms of interest will work at the track level. That is, the algorithms will work with processed detections associated over time to produce only location and time data for each vehicle. Also, since we are interested in constructing network relationships and not behavioral phenomena, we do not simulate low-level micro-behavior of traffic, such as stop lights, multiple lanes, collision avoidance, and traffic congestion. These microscopic traffic phenomena can be studied in depth with simulators such as SUMO (Simulation of Urban Mobility) \cite{sumo}.
	
		\subsubsection*{Adapting the Activity Model}

			We create an observational model for the modality of human mobility, specifically people driving throughout a city. To adapt the activity model to this application we let \emph{agents} be people within a city. Then each agent performs the \emph{action} of driving to a specific type of location dictated by the agent's \emph{role} at that time.
		
		\subsubsection*{Creating the Road Network}
		\label{subsub:creating_road_network}	

			We can define a road network for any almost any city of interest using open-source mapping data. One source of data is the OpenStreetMap (OSM) wiki \cite{osm}, which contains publicly-available maps from around the world. We are interested in two OSM data elements: \emph{nodes} and \emph{ways}. \emph{Nodes} are singular geospatial points. They are grouped together into ordered lists called \emph{ways} that define linear features such as roads or closed polygons representing areas or perimeters. Nodes and ways can also carry attributes, such as residential areas and highway types.
	
\begin{figure}[ht]
				\centering
				\includegraphics[width=1in]{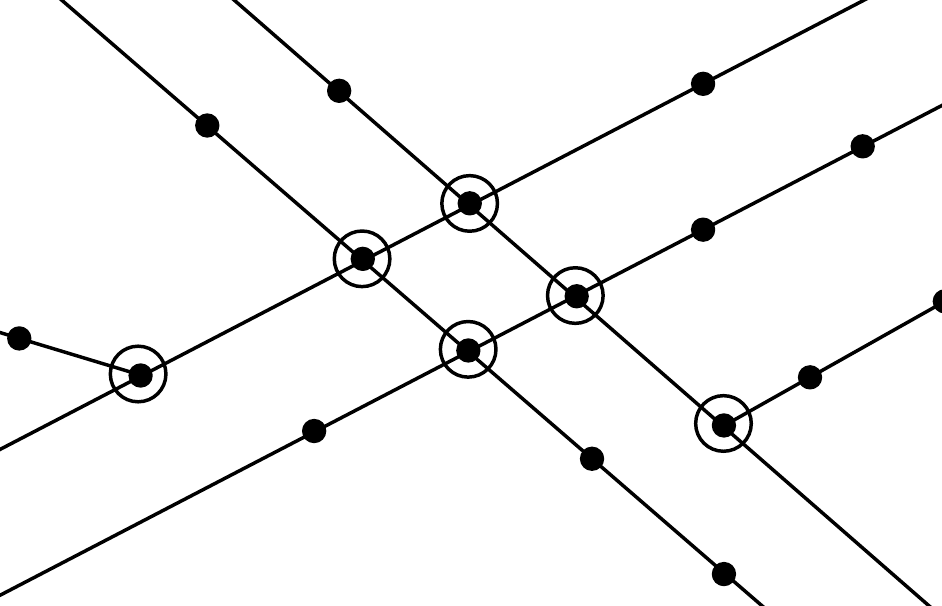}
				\includegraphics[width=1in]{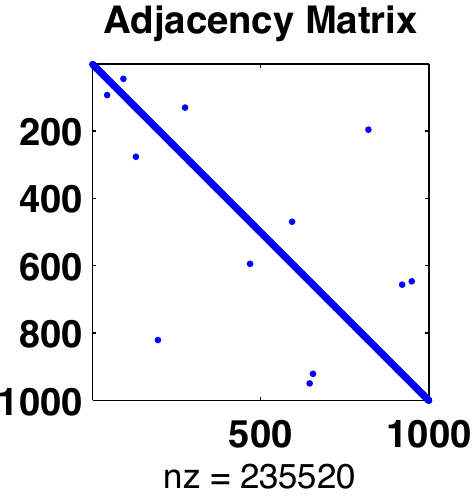}
				\caption{Illustration of road network representation. Dots are nodes along the road and circles are intersections. A link exists for nodes along a common road and for nodes that are intersections. Distances are labeled between nodes. These relationships are stored in a weighted adjacency matrix.}
				\label{fig:road_network}
			\end{figure}

			We represent the road network with a graph where vertices are points along every road. An edge exists between two points if they are adjacent on the same road or if they coincide to indicate an intersection. The graph is stored as a weighted adjacency matrix whose weights correspond to the physical length of the node-to-node road segments. Thus, we also have an encoding of the physical distances between each node. This is illustrated in Figure \ref{fig:road_network}.

		\subsubsection*{Physical Location Assignment}
		\label{subsub:road-network}
	
			Next, we label the nodes in the road network according to the types of physical locations they represent. There are two approaches to this labeling depending on the types of map data that are available.
	
			For the first case, node labeling is done directly. This is possible with vectorized data where a one-to-one labeling is made for each node. For example, each node and way in the OSM data format can be attributed to indicate characteristics like businesses or highway types.
	
			For the other case, node labeling is done indirectly. Either a vectorized set of points defines a polygonal area encompassing multiple nodes, or rasterized data specifies a grid whose individual regions may cover several nodes at once. Labeling can be done by assigning all nodes within a region with the same attributes.
	
			Depending on the application, labels may not be available for each type of location we would like to simulate. In this case we can infer node types based on related categories. For example, land usage data is available from the National Land Cover Database (NLCD) published by the U.S. Geological Survey \cite{nlcd}. The NLCD data comes in raster format and enumerates each pixel with Census Feature Class Codes (CFCC) that describes different levels of urban development as well as housing and commercial zones. Therefore we can label each node in the road network with a CFCC and then map that label to a related location type. For example, in areas of high urban development, we could label nodes as high-density apartments or small businesses. Because a CFCC could encapsulate multiple types of roles we define a probability vector $\pi_c$ for each CFCC, which is a $1 \times R$-dimensional vector. Then for each node we draw from $\sim Multinomial(\pi_c)$ to determine which role that node adopts for the simulation. In this way we will end up with neighborhoods of generally similar locations but still maintain diversity.	

		\subsubsection*{Creating Vehicle Routes}
		\label{subsub:creating_vehicle_routes}	

			After defining the road network and labeling the nodes of interest, we can create the final vehicle tracks. To do this we find a route between any two nodes by traversing the graph of the road network using a pathfinding algorithm. While any number of algorithms can be used we assume that people always take the shortest path. In this paper we use Dijkstra's shortest path algorithm \cite{dijkstra}. However, (non-optimal) algorithms can be easily used instead, such as the breadth-first-search algorithm, A* \cite{a_star}, or for extremely large networks, contraction hierarchies \cite{contraction}.
	
			The output of the pathfinding algorithm is a sequence of nodes. However, to produce a proper track containing vehicle positions over regular time intervals, we need to assume basic vehicle motion parameters and sensor observation parameters. In terms of vehicle motion, we sample a distribution of expected vehicle speed to assign to each track. In terms of sensor observations, we assume a zero-mean, Gaussian-distributed noise on the vehicle position and that observations are reported at a constant frame rate. The particular frame rate for each track is drawn from a distribution of expected frame rates in case the sensor reports measurements asynchronously.
	
	
	\subsection{Results} 
	\label{sub:observational_model_results}

		A model is only useful if it simulates data faithful to the target. In this section we show the results of the activity and observation models with parameters set to match the NGA data set as closely as possible and show a comparison of simulated vehicle tracks to those from the NGA data set.
	
\begin{figure}[ht]
			\centering
			\includegraphics[width=.2\textwidth]{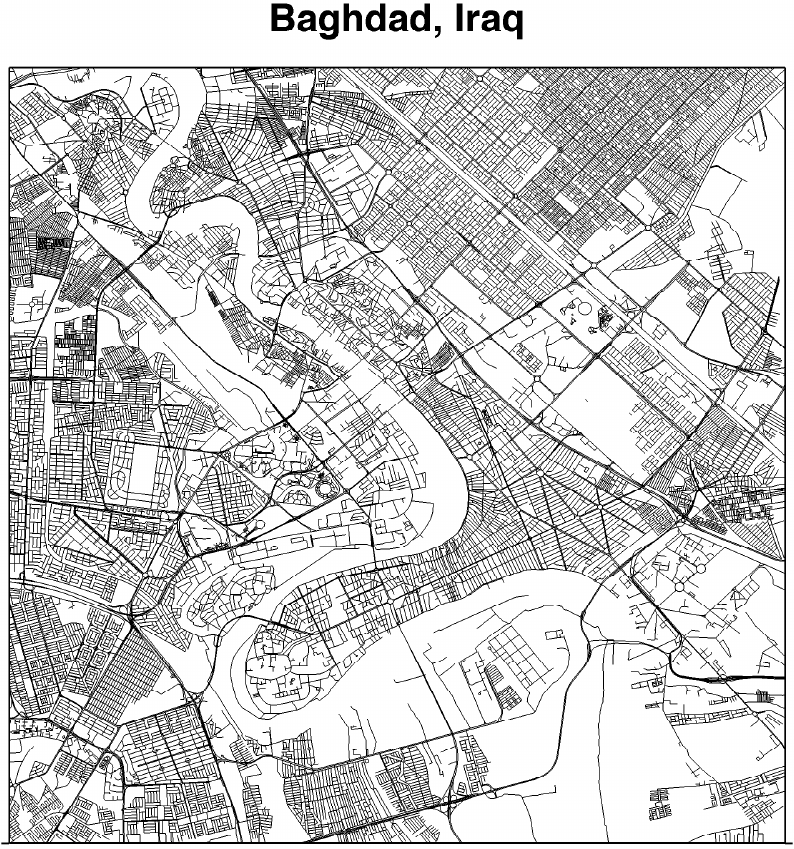}
			\caption{Full road network for Baghdad constructed with OSM data on latitude and longitude axes.}
			\label{fig:full_road_network}
		\end{figure}

		The full road network for Baghdad using OSM map data is shown in Figure \ref{fig:full_road_network}. It is comprised of 14,087 ways, 68,548 nodes, and covers approximately 10km by 10km.
	
		For comparison purposes, we used the location labels from the NGA data set to label the nodes in our road network. For each node in the NGA data having a location category, we chose the nearest-neighbor node on our road network to have the same location category.
			
		To produce our track results, we had to adjust only two parameters of the observation model: the distribution of vehicle velocity and the distribution of observation frame rate. To determine these distributions, we generated a histogram for each with 100 bins and normalized to create a proper probability mass function. Then during each phase of creating a track, these two distributions were sampled to determine the respective parameters for that track.
	
\begin{figure}[ht]
			\centering
			\includegraphics[width=1.25in]{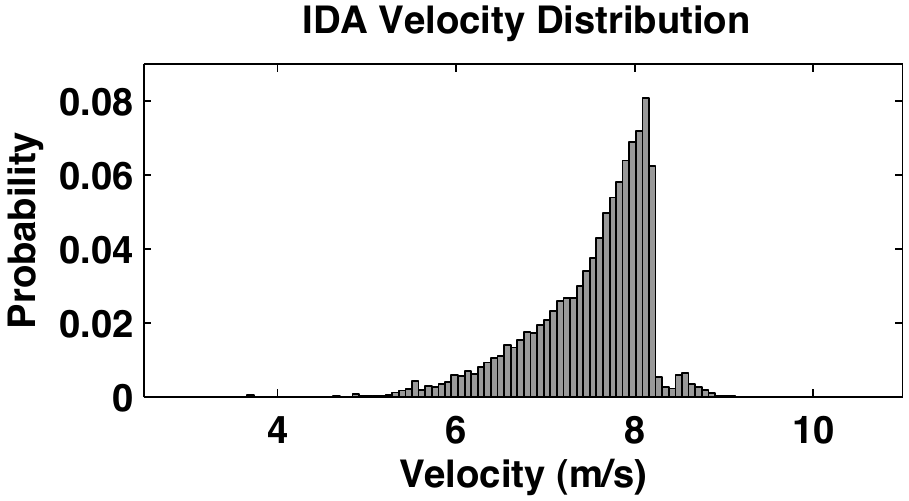}
			\includegraphics[width=1.25in]{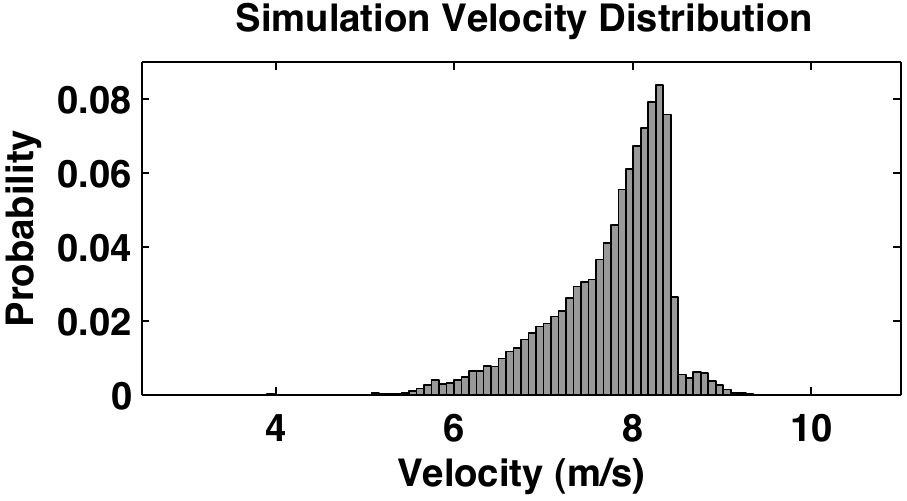}
			\includegraphics[width=1.25in]{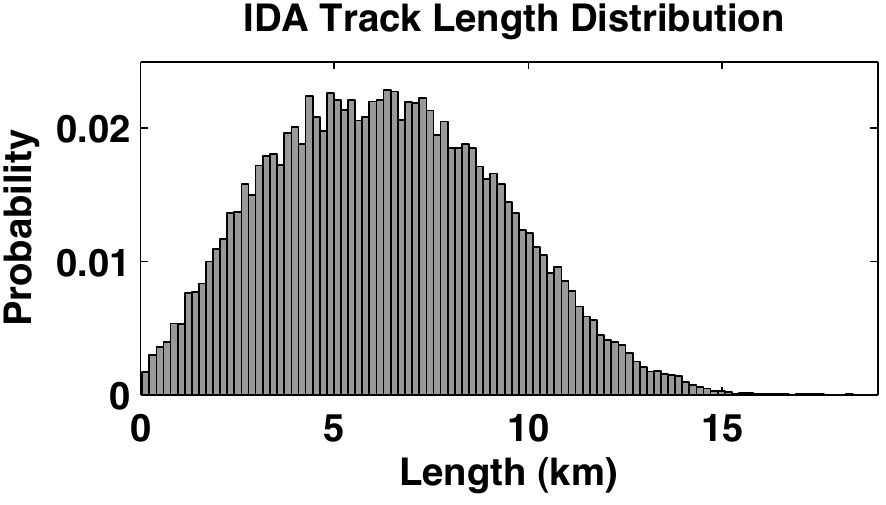}
			\includegraphics[width=1.25in]{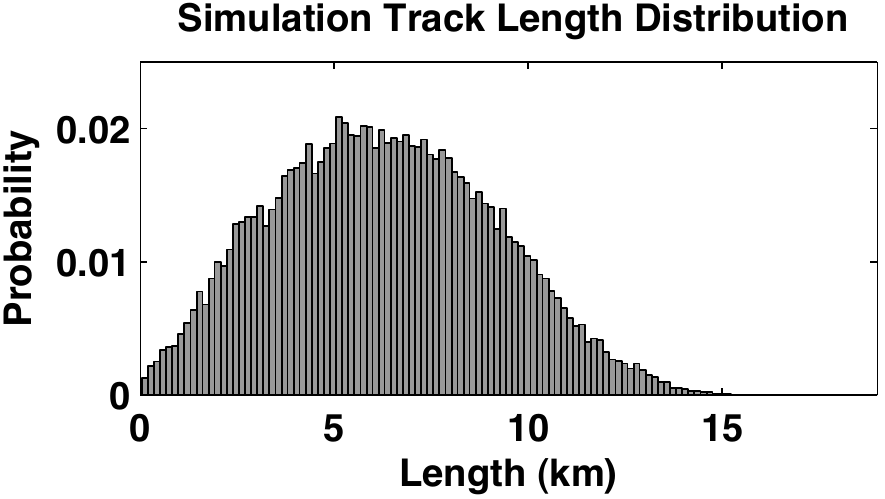}
			\caption{Comparison of NGA and simulated tracks: (a) velocity distribution and (b) track length distribution.}
			\label{fig:track_distributions}
		\end{figure}

		Figure \ref{fig:track_distributions} shows the distributions of track velocity and track length for our simulation compared to the NGA data after running the entire simulation for all agents. As expected, the velocity distribution matches very closely. An unanticipated result is the similarity of the track length distributions, which are influenced by the similar labeling of the road network locations and the shortest-path assumption of the pathfinding algorithm.
	
\begin{figure}[ht]
			\centering
			\includegraphics[width=.15\textwidth]{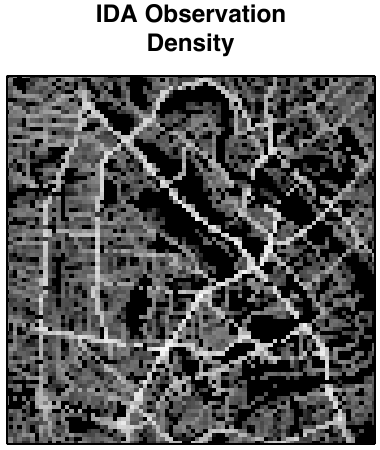}
			\includegraphics[width=.15\textwidth]{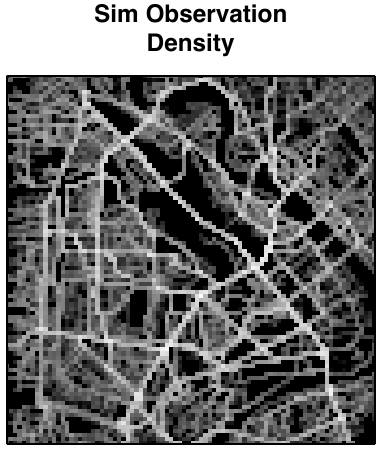}
			\caption{Comparison of NGA and simulation observation density heatmaps. Density level is on logarithmic scale.}
			\label{fig:heatmaps}
		\end{figure}

		Figure \ref{fig:heatmaps} shows measurement density heatmaps for the the NGA data and our simulation. As a result of matching the observation frame rate, the overall density of our simulation closely resembles that of NGA. Another notable similarity is that the traffic behavior in our simulation resembles NGA data very well. Major highways have the heaviest traffic while secondary roads are proportionately less dense.

\section{CONCLUSION} 
\label{sec:conclusion}	
	
		This paper presents a novel, mixed-membership, agent-based simulation model to generate network activity data for a broad range of research applications. The model combines the best of real world data, statistical simulation models, and agent-based simulation models by being easy to implement, having narrative power, and providing statistical diversity through random draws. We apply this framework to study human mobility and demonstrate the model's utility in generating high fidelity traffic data for network analytics. We also adapted the model to a high-fidelity NGA data set and showed that the model can replicate its important properties. 
		
		While we were able to closely match the average agent activity to the target NGA dataset it is clear that the activity model as is may not be rich enough to satisfy all research needs. Agents' propensity towards locations is currently defined on the population level so there is no direct sense of community, an important aspect of many network analyses. To extend the model, we will introduce the concept of lifestyles which will group agents together by similar affinities to locations. Additionally, locations in the NGA dataset are of only one type each but we foresee the power of allowing an activity to be executed while an agent is in various different roles. To do so we will apply the mixed-membership concept to actions in terms of roles and allow that to vary over time to create even greater richness. We also wish to further explore the parameterization of the activity model by adapting it to applications other than human mobility and be able to use the subsequent sensor output. To do so we will develop an observational model for a different application such as interactions on social networks such as Facebook which will also demonstrate our models' utility to different types of researchers.


{\scriptsize
\bibliographystyle{plain}
\bibliography{references}
}

\end{document}